\documentclass[10pt,letterpaper,english]{article}
\pdfoutput=1
\usepackage{mathptmx}
\usepackage{helvet}
\usepackage{courier}
\usepackage[T1]{fontenc}
\usepackage[latin9]{inputenc}
\usepackage{babel}
\usepackage{amsthm}
\usepackage{amsmath}
\usepackage[authoryear]{natbib}
\usepackage[unicode=true,pdfusetitle,
 bookmarks=false,
 breaklinks=false,pdfborder={0 0 1},backref=false,colorlinks=false]
 {hyperref}
\hypersetup{
 pdfsubject={v. Neumann's treatment of the measurement process and the collapse postulate},
 pdfkeywords={measurement process, repeatable measurement, v. Neumann measurement, collapse, reduction, projection}}

\makeatletter

\pdfpageheight\paperheight
\pdfpagewidth\paperwidth

\numberwithin{equation}{section}
\numberwithin{figure}{section}

\@ifundefined{date}{}{\date{}}
\makeatother

\begin{document}

\title{Repeatable measurements and the collapse postulate}

\author{Michael Zirpel}
\maketitle
\begin{abstract}
J. v. Neumann justified the collapse postulate by the empirical fact
of the repeatability of a measurement at a single quantum system.
However, in his quantum mechanical treatment of the measurement process
repeatability emerges without collapse. The entangled state of the
measurement device and the measured system after their interaction
ensures it already. Furthermore, this state gives the same predictions
for the measured system alone as the description demanded by the collapse
postulate.

\end{abstract}

\section{Introduction and overview}

None of the fundamental postulates of quantum mechanics is as controversial
as the collapse (reduction, projection) postulate. It demands for
a measurement an instantaneous, non-deterministic state transition,
which is seemingly in conflict with the continuous, deterministic
state evolution governed by the Schrödinger equation. In pilot wave
\citep{Bohm1952} or ensemble \citep{Ballentine1998} interpretations
this postulate is omitted, but it is part of the orthodox Copenhagen
interpretation \citep{Heisenberg1958} and of the modern quantum computing
\citep{NielsenChuang2000}.

\citet{vNeumann1932} founded his reasoning for this postulate on
the Compton-Simon experiment, where the measurement results for the
momentum of a scattered photon and the associated recoil electron
determine each other \citep{Compton1927}. He explained this ``sharp
(causal) correlation'' by the state reduction caused by the first
measurement. 

However, in v. Neumann's quantum mechanical treatment of the measurement
process the repeatability of the measurement emerges without collapse,
just by application of Born's probabilistic interpretation of the
wave function. The entangled state of the measurement device and the
measured system after their interaction ensures already, that an immediate
repetition of the process with a second measurement device of the
same type will give with probability 1 the same result as the first.
Furthermore, this entangled state of the compound system gives the
same probabilities as the description demanded by the collapse postulate
for all following measurements at the measured system.

\section{Basic notions and notations}

Let $S$ be a quantum mechanical system, described with Hilbert space
$\mathcal{H},$ and $A=\underset{k}{\sum}a_{k}\mathrm{\bigl|\mathrm{\alpha}_{\mathit{k}}\bigr\rangle}\mathrm{\bigl\langle\mathrm{\alpha}_{\mathit{k}}\bigr|}\in\mathcal{L}(\mathcal{H})$
a discrete, non-degenerate observable. For a measurement of the observable
$A$ at the system $S$ in the state $\mathrm{\bigl|\psi\bigr\rangle\in\mathcal{H}}$ 
\begin{itemize}
\item the \emph{probability postulate} (Born's rule) demands, that \emph{the
probability, to get the measurement result $a_{j}$, is $\left|\bigl\langle\mathrm{\alpha}_{j}|\psi\bigr\rangle\right|^{2}$}, 
\item the \emph{collapse postulate} demands, that, \emph{if the measurement
result is $a_{j}$, the state immediately after the measurement will
be $\bigl|\mathrm{\alpha}_{\mathit{j}}\bigr\rangle$ .} 
\end{itemize}
As a consequence of both postulates the system $S$ has to be described
after the measurement, if the result is unknown, by a mixture of states
represented by the statistical operator 
\begin{equation}
W=\underset{j}{\sum}\bigl|\bigl\langle\mathrm{\alpha}_{j}|\psi\bigr\rangle\bigr|^{2}\mathrm{\bigl|\mathrm{\alpha}_{\mathit{j}}\bigr\rangle}\mathrm{\bigl\langle\mathrm{\alpha}_{\mathit{j}}\bigr|}\label{eq:MixedState}
\end{equation}

v. Neumann's quantum mechanical treatment of the measurement process
describes a measurement\emph{ }as an interaction between the system
$S$ and a measurement device $M$, which is itself a quantum system
with Hilbert space $\mathcal{H}_{M}$. Pairwise orthogonal pointer
states $\bigl|\mathrm{\varphi}_{k}^{(M)}\bigr\rangle\in\mathcal{H}_{M}$
indicate the measurement results. The interaction of the measured
system and the measurement device is described by an unitary transformation
$U^{(SM)}$ in the tensor product Hilbert space $\mathcal{H}\otimes\mathcal{H}_{M}$
of the compound system $SM$. The assumption, that an \emph{ideal
measurement} of an eigenstate $\bigl|\mathrm{\alpha}_{\mathit{k}}\bigr\rangle$
of the measured observable $A$ should give exactly the corresponding
pointer state $\bigl|\varphi_{k}^{(M)}\bigr\rangle$ as result, without
disturbing the system $S$, can be expressed by

\begin{equation}
U^{(SM)}\bigl|\mathrm{\alpha}_{k}\bigr\rangle\bigl|\mathrm{\varphi}_{0}^{(M)}\bigr\rangle=\bigl|\mathrm{\alpha}_{k}\bigr\rangle\bigl|\mathrm{\varphi}_{k}^{(M)}\bigr\rangle\label{eq:CondvNeumann}
\end{equation}
Therefore, with the initial state $\mathrm{\bigl|\psi\bigr\rangle}$
of the system $S$ the unitary transformation $U^{(SM)}$ will give
the entangled final state
\[
\mathrm{\bigl|\Phi\bigr\rangle}=U^{(SM)}\mathrm{\mathrm{\bigl|\psi\bigr\rangle}\bigl|\mathrm{\varphi_{0}}^{(\mathit{M})}\bigr\rangle}=\underset{k}{\sum}\bigl\langle\mathrm{\alpha}_{k}|\psi\bigr\rangle\bigl|\mathrm{\alpha}_{k}\bigr\rangle\bigl|\mathrm{\varphi}_{k}^{(M)}\bigr\rangle
\]
The reading of the pointer can be considered as a secondary measurement.
The probability of the result $A_{j}^{(M)}=1\otimes\bigl|\mathrm{\varphi}_{j}^{(M)}\bigr\rangle\bigl\langle\mathrm{\varphi}_{j}^{(M)}\bigr|$,
that the pointer state $\bigl|\mathrm{\varphi}_{j}^{(M)}\bigr\rangle$
is observed, is according the probability postulate for the final
state $\mathrm{\bigl|\Phi\bigr\rangle}$ given by 
\begin{equation}
p(A_{j}^{(M)})=\bigl\langle\Phi\,\bigl|\mathrm{\varphi}_{j}^{(M)}\bigr\rangle\bigl\langle\mathrm{\varphi}_{j}^{(M)}\bigl|\,\Phi\bigr\rangle=\bigl|\bigl\langle\mathrm{\alpha}_{j}|\psi\bigr\rangle\bigr|^{2}\label{eq:ResultProbability}
\end{equation}

As long as the compound system $SM$ is in the state $\mathrm{\bigl|\Phi\bigr\rangle}$,
the system $S$ alone has to be described by the statistical operator
given by the partial trace
\begin{equation}
\textrm{tr}_{\mathcal{H}_{M}}(\bigl|\Phi\bigr\rangle\bigl\langle\Phi\bigl|)=\underset{k}{\sum}\bigl\langle\mathrm{\alpha}_{k}|\psi\bigr\rangle\mathrm{\bigl\langle\psi|\mathrm{\alpha}_{\mathit{k}}\bigr\rangle\bigl|\mathrm{\alpha}_{\mathit{k}}\bigr\rangle}\mathrm{\bigl\langle\mathrm{\alpha}_{\mathit{k}}\bigr|}=W\label{eq:PartialTrace}
\end{equation}
which is identical with (\ref{eq:MixedState}). But this description
gives no statement about the correlation with the measurement result.

\section{Repeatability of the measurement}

The wave function of the compound system can be used to compute the
conditional probabilities of the results of further measurements.
When a second measurement device $M'$ of the same type interacts
with the measured system $S$ as part of the compound system $SM$
in state $\bigl|\Phi\bigr\rangle$ in the enlarged Hilbert space $\mathcal{H}\otimes\mathcal{H}_{M}\otimes\mathcal{H}_{M'}$,
this gives the state
\[
\bigl|\mathrm{\Phi'}\bigr\rangle=U^{(SM')}\bigl|\Phi\bigr\rangle\bigl|\mathrm{\varphi_{0}}^{(M')}\bigr\rangle=\underset{k}{\sum}\bigl\langle\mathrm{\alpha}_{k}|\psi\bigr\rangle\bigl|\mathrm{\alpha}_{k}\bigr\rangle\bigl|\mathrm{\varphi}_{k}^{(M)}\bigr\rangle\bigl|\mathrm{\varphi}_{k}^{(M')}\bigr\rangle
\]
The pointer results of both measurement devices define a Boolean event
algebra and a common probability space, because all projections onto
the pointer states $A_{j}^{(M)}=1\otimes\bigl|\mathrm{\varphi}_{j}^{(M)}\bigr\rangle\bigl\langle\mathrm{\varphi}_{j}^{(M)}\bigr|\otimes1^{(M')}$
and $A_{k}^{(M')}=1\otimes1^{(M)}\otimes\bigl|\mathrm{\varphi}_{k}^{(M')}\bigr\rangle\bigl\langle\mathrm{\varphi}_{k}^{(M')}\bigr|$
commute pairwise. Therefore, the conditional probability to get the
pointer result $A_{k}^{(M')}$ in the second measurement, given the
pointer result $A_{j}^{(M)}$ in the first, is well-defined 
\begin{equation}
p(\mathrm{\mathit{A}}_{k}^{(M')}|\,\mathit{A}_{j}^{(M)})=\frac{p(\mathrm{\mathit{A}}_{k}^{(M')}\wedge\mathit{A}_{j}^{(M)})}{p(A_{j}^{(M)})}=\frac{\bigl\langle\Phi'\,\bigl|\mathrm{\varphi}_{j}^{(M)}\bigr\rangle\bigl\langle\mathrm{\varphi}_{j}^{(M)}\bigl|\otimes\bigr|\mathrm{\varphi}_{k}^{(M')}\bigr\rangle\bigl\langle\mathrm{\varphi}_{k}^{(M')}\bigl|\,\Phi'\bigr\rangle}{\bigl\langle\Phi'\,\bigl|\mathrm{\varphi}_{j}^{(M)}\bigr\rangle\bigl\langle\mathrm{\varphi}_{j}^{(M)}\bigl|\,\Phi'\bigr\rangle}=\delta_{j,k}\label{eq:CondProbRep}
\end{equation}
This means: \emph{The probability is $1$, that an immediate repetition
of the measurement gives the same result as the first.} The same statement
is valid for further repetitions%
\footnote{This gives a simple explanation of avalanche effects, which are part
of some measurement devices, where many molecules interact as measurement
devices with a measured particle and the resulting state is macroscopic
visible, because all molecules are observed in the same pointer state.%
} and, when the initial state of the system is a mixture or when the
measured observable is degenerate%
\footnote{But the spectrum of the observable has to be discrete \citep{BuschLahtiMittelstaedt1991}.%
}. 

This repeatability of the measurement is a consequence of condition
(\ref{eq:CondvNeumann}). The weaker condition 
\[
U^{(SM)}\mathrm{\bigl|\mathrm{\alpha}_{\mathit{k}}\bigr\rangle\bigl|\mathrm{\varphi_{0}}^{(\mathit{M})}\bigr\rangle}=\bigl|\mathrm{\psi}_{k}\bigr\rangle\bigl|\mathrm{\varphi}_{k}^{(M)}\bigr\rangle
\]

\noindent with some not further specified $\bigl|\mathrm{\psi}_{k}\bigr\rangle\in\mathcal{H}$,
gives the same probability distribution for the pointer results (\ref{eq:ResultProbability}),
but in general without repeatability.

\section{Description of the measured system}

With v. Neumann's description of a repeatable measurement all further
measurements at the measured system have without collapse the same
conditional probability distributions as demanded by the collapse
postulate.\emph{ }To see that, let instead of $M'$ another device
$M''$ for the measurement of an arbitrary non-degenerate observable
$B=\underset{k}{\sum}b_{k}\mathrm{\bigl|\mathrm{\beta}_{\mathit{k}}\bigr\rangle}\mathrm{\bigl\langle\mathrm{\beta}_{\mathit{k}}\bigr|\in\mathcal{L}(\mathcal{H})}$
interact with the measured system $S$ as part of the compound system
$SM$ in state $\bigl|\Phi\bigr\rangle$. With the resulting state
\[
\bigl|\mathrm{\Phi''}\bigr\rangle=\underset{k,j}{\sum}\bigl\langle\mathrm{\alpha}_{k}|\psi\bigr\rangle\bigl\langle\mathrm{\beta}_{j}|\mathrm{\alpha}_{k}\bigr\rangle\bigl|\mathrm{\beta}_{j}\bigr\rangle\bigl|\mathrm{\varphi}_{k}^{(M)}\bigr\rangle\bigl|\mathrm{\varphi}_{j}^{(M'')}\bigr\rangle
\]
the conditional probability, to get in the second measurement the
pointer result $\mathrm{\mathit{B}}_{k}^{(M'')}=1\otimes1^{(M)}\otimes\bigl|\mathrm{\varphi}_{j}^{(M'')}\bigr\rangle\bigl\langle\mathrm{\varphi}_{j}^{(M'')}\bigr|$,
if the first measurement result is $\mathit{A}_{j}^{(M)},$ is also
well-defined 
\[
p(\mathrm{\mathit{B}}_{k}^{(M'')}|\,\mathit{A}_{j}^{(M)})=\frac{\bigl\langle\Phi''\bigl|\,\mathrm{\varphi}_{j}^{(M)}\bigr\rangle\bigl\langle\mathrm{\varphi}_{j}^{(M)}\bigl|\otimes\bigr|\mathrm{\varphi}_{k}^{(M'')}\bigr\rangle\bigl\langle\mathrm{\varphi}_{k}^{(M'')}\bigl|\,\Phi''\bigr\rangle}{\bigl\langle\Phi''\,\bigl|\mathrm{\varphi}_{j}^{(M)}\bigr\rangle\bigl\langle\mathrm{\varphi}_{j}^{(M)}\bigl|\,\Phi''\bigr\rangle}=\bigl|\bigl\langle\mathrm{\beta}_{k}|\mathrm{\alpha}_{j}\bigr\rangle\bigr|^{2}
\]
This exactly the same value as with the collapse postulate; (\ref{eq:CondProbRep})
is just a special case with $B=A$. 

The total probability of the result $\mathrm{\mathit{B}}_{k}^{(M'')}$
\[
p(B_{k}^{(M'')})=\sum_{j}p(\mathrm{\mathit{B}}_{k}^{(M'')}|\,\mathit{A}_{j}^{(M)})p(A_{j}^{(M)})=
\]
\[
\sum_{j}\bigl|\bigl\langle\mathrm{\alpha}_{j}|\psi\bigr\rangle\bigr|^{2}\bigl\langle\mathrm{\beta}_{k}\bigl|\mathrm{\alpha}_{j}\bigr\rangle\bigl\langle\mathrm{\alpha}_{j}\bigl|\mathrm{\beta}_{k}\bigr\rangle=\textrm{tr}(W\bigl|\mathrm{\beta}_{k}\bigr\rangle\bigl\langle\mathrm{\beta}_{k}\bigl|)
\]
is the same as the probability of the result $b_{k}$ for the mixture
$W$ in (\ref{eq:MixedState}) and (\ref{eq:PartialTrace}).

So far we have considered only the situation immediately after the
measurement $M$. With the collapse postulate it is possible to describe
the system $S$ by a pure state $\bigl|\mathrm{\alpha}_{j}\bigr\rangle\in\mathcal{H}$,
whose evolution is governed by a group of unitary transformations
$U_{t}\in\mathcal{L}(\mathcal{H})$, as long as $S$ is isolated.
If the measurement result was $a_{j}$, the state at time $t$ after
the measurement will be 
\[
\bigl|\mathrm{\alpha}_{j}(t)\bigr\rangle=U_{t}\bigl|\mathrm{\alpha}_{j}\bigr\rangle
\]

If $S$ is isolated and the pointer state $\bigl|\mathrm{\varphi}_{k}^{(M)}\bigr\rangle$
is an eigenstate of the evolution of $M$, the state of the compound
system $SM$ at the time $t$ after the measurement will be
\[
\mathrm{\bigl|\Phi_{\mathit{t}}\bigr\rangle}=\sum_{k}\bigl\langle\mathrm{\alpha}_{k}|\psi\bigr\rangle\bigl|\mathrm{\alpha}_{k}(t)\bigr\rangle\bigl|\mathrm{\varphi}_{k}^{(M)}\bigr\rangle
\]
With this state the conditional probability, to get in a second measurement
$M_{t}''$ at time $t$ the result $\mathrm{\mathit{B}}_{k}^{(M_{t}'')}$,
if the first measurement result at time $t=0$ was $\mathit{A}_{j}^{(M)}$,
is
\begin{equation}
p(\mathrm{\mathit{B}}_{k}^{(M_{t}'')}|\,\mathit{A}_{j}^{(M)})==\frac{\bigl\langle\Phi_{\mathit{t}}\,\bigl|\mathrm{\varphi}_{j}^{(M)}\bigr\rangle\bigl\langle\mathrm{\varphi}_{j}^{(M)}\bigl|\otimes\bigr|\mathrm{\varphi}_{k}^{(M_{t}'')}\bigr\rangle\bigl\langle\mathrm{\varphi}_{k}^{(M_{t}'')}\bigl|\,\Phi_{\mathit{t}}\bigr\rangle}{\bigl\langle\Phi_{\mathit{t}}\,\bigl|\mathrm{\varphi}_{j}^{(M)}\bigr\rangle\bigl\langle\mathrm{\varphi}_{j}^{(M)}\bigl|\,\Phi_{\mathit{t}}\bigr\rangle}=\bigl|\bigl\langle\mathrm{\beta}_{k}|\mathrm{\alpha}_{j}(t)\bigr\rangle\bigr|^{2}\label{eq:DynCondProp}
\end{equation}
This is the same value as with the collapse postulate. 

Interactions of other systems with the measurement device $M$ can
destroy the measurement result and change the state of the system
$S$ remotely. It is an open question how a measurement result becomes
irreversible. However, pointer readings by a repeatable measurement
at $M$ with another device $\widetilde{M}$ will give a state 

\[
\mathrm{\bigl|\widetilde{\Phi}_{\mathit{t}}\bigr\rangle}=\sum_{k}\bigl\langle\mathrm{\alpha}_{k}|\psi\bigr\rangle\bigl|\mathrm{\alpha}_{k}(t)\bigr\rangle\bigl|\mathrm{\varphi}_{k}^{(M)}\bigr\rangle\bigl|\mathrm{\widetilde{\varphi}}_{k}^{(\widetilde{M})}\bigr\rangle
\]
which does not change the conditional probabilities for the results
of a further measurement $M_{t}''$ (\ref{eq:DynCondProp}). A similar
mechanism is assumed by decoherence theory \citep{Schlosshauer2004}.

\section{Conclusions}

Applying the probability postulate we have shown that v. Neumann's
treatment of the measurement process describes repeatable measurements
without collapse of the wave function. However, this description has
to take into account not only the measured system but also the measurement
devices. A simpler description of the measured system alone is possible,
which gives the same conditional and total probabilities for all further
measurement results: It is the description demanded by the collapse
postulate.

One can regard this as a derivation of the collapse postulate, without
contradicting impossibility proofs like (\citealp{BassiGhirardi2000}).
From this point of view the wave function is merely a computational
tool. And the collapse is no physical process; it is only the change
to an equivalent but simpler probabilistic description with a \emph{cut}
between the observed system and the rest of the world. The dynamical
conditions during and after the measurement interaction determine,
if this cut is possible. 

Of course, our analysis does not explain how definite measurement
results arise. That is already presupposed by the probability postulate.
Nevertheless, it explains why the collapse of the wave function gives
a correct probabilistic description of the measured system, whenever
a definite measurement result was obtained by a repeatable measurement.
And it explains how it is possible to omit the collapse postulate
at all, without losing the capability to describe sequential measurements.
\bibliographystyle{apsrev}

\end{document}